\documentclass[aps,prd,showpacs]{revtex4}

\usepackage{amsmath,amsfonts}
\usepackage[dvips]{graphicx}

\newcommand{\tr}{\mathop{\text{tr}}}
\newcommand{\ud}{\mathrm{d}}

\begin{document}

\title{Paths of Friedmann-Robertson-Walker brane models} 
\author{Marek Szyd{\l}owski}
\email{uoszydlo@cyf-kr.edu.pl}
\affiliation{Astronomical Observatory, Jagiellonian University, Orla 171, 30-244
Krak{\'o}w, Poland}
\affiliation{Marc Kac Complex Systems Research Center, Jagiellonian University, 
Reymonta 4, 30-059 Krak{\'o}w, Poland}
\author{Orest Hrycyna}
\email{hrycyna@kul.lublin.pl}
\affiliation{Department of Theoretical Physics, Faculty of Philosophy, The John
Paul II Catholic University of Lublin, Al. Rac{\l}awickie 14, 20-950 Lublin,
Poland}

\date{\today}

\begin{abstract}
Dynamics of brane-world models of dark energy is reviewed. We demonstrate that
simple dark energy brane models can be represented as 2-dimensional dynamical 
systems of a Newtonian type. Hence a fictitious particle moving in a potential 
well characterizes the model. We investigate the dynamics of the brane models 
using methods of dynamical systems. The simple brane-world models can be 
successfully unified within a single scheme -- an ensemble of brane dark energy 
models. We characterize generic models of this ensemble as well as exceptional 
ones using the notion of structural stability (instability). Then due to the 
Peixoto theorem we can characterize the class of generic brane models. We show 
that global dynamics of the generic brane models of dark energy is topologically 
equivalent to the concordance $\Lambda$CDM model. We also demonstrate that the 
bouncing models or models in which acceleration of the universe is only 
transient phenomenon are non-generic (or exceptional cases) in the ensemble. We 
argue that the adequate brane model of dark energy should be a generic case in 
the ensemble of FRW dynamical systems on the plane.
\end{abstract}


\maketitle

\section{Introduction}

Recent observations of distant supernovae type Ia (SNIa)
\cite{Perlmutter:1999,Riess:1998} as well as other current observations of 
cosmic microwave background radiation (CMBR)
anisotropies indicate that our universe is almost flat and stay in the
accelerated phase of expansion \cite{Bernardis:2000,Lange:2001}. Principally
there are two alternatives of explanation properties of current universe. In the
first approach the universe is accelerating due to presence of mysterious energy
called dark energy $X$, of unknown origin, which violates the strong energy condition $\rho_{X}+3 p_{X} > 0$, where $\rho_{X}$ and $p_{X}$ are energy density and pressure of dark energy, respectively. Alternative to the dark energy idea is the explanation of acceleration of the universe based on some modification to the Friedmann-Robertson-Walker (FRW) dynamics arising from new physics (i.e., extra dimensions). In this approach dark energy is the manifestation of modified gravitational dynamics of the $4$-dimensional brane leading to a self-accelerated universe. In this conception which is called brane dark energy new physics mimic on the phenomenological level dark energy or effects of extra dimensions manifest themselves as a modification to the Friedmann equation which govern the evolution of the $4$-dimensional brane localized in the higher-dimensional bulk space. In contrast to the classical Kaluza-Klein theories (when extra dimensions, which manifest themselves at high energies, are either compact and have finite volume), theories with infinite volume of extra dimensions modify the late time cosmological evolution of the universe and gravitational dynamics is preserved at short distances.

The hierarchy problem \cite{Arkani:1998,Randall:1999a,Randall:1999b} and the problem of present acceleration of the Universe \cite{Peebles:2003,Turner:2000} also motivate all theories with large extra dimensions. In the literature of the subject there are many different propositions of dark energy of brane origin (for review see \cite{Sahni:2005}).
Different brane-world scenarios were tested by recent astronomical measurements
of distant supernovae Ia~\cite{Avelino:2002} as well as other observations from
Type IIb radio galaxies and X-ray gas mass fraction~\cite{Zhu:2004,Zhu:2002}. The
stringent constraint on the model parameters can be also obtained from baryon
oscillation peak measurements~\cite{Firbairn:2006,Alam:2006,Szydlowski:2006}.

The main aim of this paper is to investigate dynamics of the most popular brane
dark energy models using methods of dynamical systems. We investigate a class of
the simplest brane models in which our observable universe is a
$(3+1)$-dimensional brane and a $3$-space is homogeneous and isotropic. The
physical space is embedded in a $(4+1)$-dimensional space called a bulk space.
For all these models we obtain an analog of the classical Friedmann equation
with some additional terms which are functions of both the scale factor $a$ and
energy density of matter on the brane (see Table \ref{tab:1}). If we assume that
the energy density satisfies the conservation condition $\dot{\rho}=-3 H (\rho
+p)$, where $p=w\rho$ is the equation of state, then $\rho=\rho(a)$ and the
modified FRW equation on the brane can be represented by 
\begin{equation}
H^{2} + \frac{k}{a^{2}} \equiv \frac{\rho_{\textrm{eff}}}{3},
\label{eq:1}
\end{equation}
where $H=(\ln a)\dot{}$ is Hubble's function, $k$ is the curvature index, $\rho_{\textrm{eff}}=\rho_{\textrm{eff}}(a)$ is the effective energy density. For the concordance $\Lambda$CDM model $\rho_{\textrm{eff}}=\rho_{\textrm{m},0}(a/a_{0})^{-3} + \Lambda$, $p_{\textrm{eff}}= 0 - \Lambda$. Without lost of degree of generality one can choose $a_{0}=1$ -- as the present value of the scale factor; $1+z=a_{0}/a$, where $z$ is redshift.

\begin{turnpage}
\begin{table}
\caption{Classification of simple dark energy brane models in terms of $H(x)$
relation and $V_{\textrm{brane}}=-\frac{\rho_{\textrm{brane}}a^{2}}{6}=-\frac{1}{2}\Omega_{\textrm{brane}}(a)a^{2}$.}
\label{tab:1}
\begin{tabular}{|p{0.7cm}|p{4.5cm}|p{8cm}|p{9cm}|}
\hline
case & model & $(\frac{H}{H_{0}})^{2}=f(x)$ & $V_{\textrm{brane}}$ \\
\hline
\hline
I & Dvali-Turner model \cite{Dvali:2003} &
$(1-\Omega_{\textrm{m},0})(\frac{H}{H_{0}})^{\alpha}+\Omega_{\textrm{m},0}x^{-3}$ & \\
II & Deffayet-Dvali-Gabadadze model $p=w \rho$ \cite{Deffayet:2002} & $\Omega_{k,0}x^{-2} +
\bigg(\sqrt{\Omega_{r_{c}}} + \sqrt{\Omega_{r_{c}} +
\Omega_{\textrm{m},0}x^{-3(1+w)}}\bigg)^{2}$
& $-\frac{1}{2}\bigg(\sqrt{\Omega_{rc}} + \sqrt{\Omega_{rc} +
\Omega_{m,0}x^{-3(1+w)}}\bigg)^{2}x^{2}$ \\
III & Shtanov-Sahni model $w=0,1/3$, $\epsilon=\pm 1$ \cite{Shtanov:2003} &
$\Omega_{\Lambda,0}+\Omega_{\textrm{m},0}x^{-3} + \Omega_{dr,0}x^{-4} +
\epsilon|\Omega_{\lambda,0}|x^{-6(1+w)}$ &
$-\frac{1}{2}\bigg(\Omega_{\Lambda,0}+\Omega_{m,0}x^{-3} + \Omega_{dr,0}x^{-4} +
\epsilon|\Omega_{\lambda,0}|x^{-6(1+w)}\bigg)x^{2}$ \\
IV & Sahni-Shtanov Brane I $(+)$ and Brane II $(-)$ models \cite{Sahni:2003}&
$\Omega_{\textrm{m},0}x^{-3} + \Omega_{\sigma} + 2\Omega_{l} \pm
2\sqrt{\Omega_{l}}\sqrt{\Omega_{\textrm{m},0}x^{-3} + \Omega_{\sigma} + \Omega_{l} +
\Omega_{\Lambda_{b}}}$ & $-\frac{1}{2}\bigg(\Omega_{m,0}x^{-3} + \Omega_{\sigma} +
2\Omega_{l} \pm
2\sqrt{\Omega_{l}}\sqrt{\Omega_{\textrm{m},0}x^{-3} + \Omega_{\sigma} + \Omega_{l} +
\Omega_{\Lambda_{b}}}\bigg)x^{2}$ \\
\hline
\end{tabular}
\end{table}
\end{turnpage}

Equation (\ref{eq:1}) can be rewritten to the new form analogous to the first integral of the energy for a unit mass particle moving in the potential well
\begin{equation}
\frac{\dot{a}^{2}}{2} + V(a) \equiv 0,
\label{eq:2}
\end{equation}
where a dot denotes the differentiation with respect to the rescaled time $t \to
\tau \colon |H_{0}| \ud t = \ud \tau$ and
$V(a)=-\frac{1}{2}\{\Omega_{\textrm{m},0}a^{-1}+\Omega_{k,0}+\Omega_{\Lambda,0}a^{2}\}$; $\Omega_{\text{m},0}=\frac{\rho_{\textrm{m},0}}{3H_{0}^{2}}$, $\Omega_{\Lambda,0}=\frac{\Lambda}{3H_{0}^{2}}$ are density parameters for matter and the cosmological constant on the brane, $V(a)=-\frac{\rho_{\textrm{eff}}a^{2}}{6}$.

Equation (\ref{eq:2}) reduces the problem of dynamics of a cosmological model to the standard problem of classical mechanics -- motion of a particle of a unit mass in the potential well $V(a)$ on the distinguished ``zero energy level''.

For the FRW brane cosmological models there is a counterpart of the Friedmann first integral (\ref{eq:2}) in the form
\begin{equation}
\frac{\dot{a}^{2}}{2} + V(a) + V_{\textrm{brane}}(a) = 0,
\label{eq:3}
\end{equation}
where $V_{\textrm{brane}}(a)=-\frac{\rho_{\text{brane}}a^{2}}{6}$, $V(a)=-\frac{\rho_{\textrm{eff}}a^{2}}{6}$.

Both relations (\ref{eq:2}) and (\ref{eq:3}) play the role of a first integral of motion for the system
\begin{equation}
\begin{array}{l}
\dot{x} = y, \\
\dot{y} = - \frac{\partial V}{\partial x},
\end{array}
\label{eq:4}
\end{equation}
which has the form of a $2$-dimensional dynamical system of a Newtonian type and $x=a/a_{0}$.

The classical motion of the system is restricted to the domain admissible for motion:
\begin{equation}
\mathcal{D}_{0} = \{ x \colon V \le 0 \}.
\label{eq:5}
\end{equation}
The evolution of the system has been visualized in $2$-dimensional phase space $(x,\dot{x})$. There are two types of solutions of (\ref{eq:4}): $1)$ a singular one which corresponds to the critical points of system~(\ref{eq:4}) such that $y_{0}=0$, and $(\partial V/\partial x)_{x_{0}}=0$; $2)$ a nonsingular one which lies on the algebraic curves determined by the Friedmann first integral $y^{2}/2 + V(x) =0$. It is convenient to extract the curvature term from the potential function. Then we can regard evolution of the system on a constant energy level $E = \frac{1}{2} \Omega_{k,0}$.

An advantage of dynamical system methods is a investigation of all admissible
solutions for all initial conditions. A picture of all evolutional paths
represented by phase curves and critical points form phase portrait -- global
visualization of dynamics. The trajectory of the flat model $\Omega_{k,0}=0$
divides phase portrait in to two subsets occupied by open models
$(\Omega_{k,0}>0)$ and by closed models $(\Omega_{k,0}<0)$.

The character (type) of the critical points is, following the Hartman-Grobman theorem, determined from the linearized system
\begin{equation}
\begin{array}{l}
(x - x_{0})\dot{} = y, \\
(y - 0)\dot{} = -(\frac{\partial^{2} V}{\partial x^{2}})_{x=x_{0}} (x - x_{0}),
\end{array}
\label{eq:6}
\end{equation}
around the critical point (which is a static critical point in any case in a finite domain of the phase space). Therefore the characteristic equation of the linearization matrix is 
\begin{equation}
\lambda^{2} + \det A = 0,
\label{eq:7}
\end{equation}
where $\tr A = 0 $, $\det A = V_{xx}(x_{0})$.

From equation~(\ref{eq:7}) we obtain that only two types of critical points are admissible for the system of a Newtonian type, namely saddles if $V_{xx}(x_{0}) < 0$ or centers in opposite case of $V_{xx}(x_{0})>0$. The case of $V_{xx}(x_{0})=0$ (a inflection point) is a degenerated case. In the first case eigenvalues of the linearization matrix are real of opposite sings. In the case of centers (non-hyperbolic critical points) eigenvalues are purely imaginary.

For the conservative system it is useful to develop methods of qualitative
investigations of differential equations \cite{Szydlowski:2006a}. The main aim of
this approach is to construct phase portraits of the system which contain global
information about the dynamics. The phase space $(x,y)$ offers a possibility of
natural geometrization of the dynamical behavior. It is in a simple $2$-dimensional case
structuralized by critical points or non-point closed trajectories (limit
cycles) and trajectories joining them. Two phase portraits are equivalent modulo
homeomorphism preserving orientation of the phase curves (or phase
trajectories). From the physical point of view critical points (and limit
cycles) represent asymptotic states (equilibria) of the system.
Equivalently, one can look at the phase flow as a vector field
\begin{equation}
\boldsymbol{f}=\bigg[y,-\frac{\partial V}{\partial x}\bigg]^{T},
\label{eq:8}
\end{equation}
whose integral curves are the phase curves.

Thanks to Andronov and Pontryagin \cite{Andronov:1937sg}, the important idea of structural
stability was introduced into the ensemble~\cite{Ellis:2004} of all dynamical systems. A vector
field, say $\boldsymbol{f}$, is a structurally stable vector field if there is
an $\varepsilon > 0$ such that for all other vector fields $\boldsymbol{g}$,
which are close to $\boldsymbol{f}$ (in some metric sense)
$\|\boldsymbol{f}-\boldsymbol{g}\|<\varepsilon$, $\boldsymbol{f}$ and
$\boldsymbol{g}$ are topologically equivalent. The notion of structural
stability is mathematical formalization of intuition that physically realistic
models in applications should posse some kind of stability, therefore small
changes of the r.h.s. of the system (i.e. vector field) doesn't disturb the
phase portrait. For example motion of pendulum is structurally unstable because
small changes of vector field (constructed from r.h.s. of the system) of a
friction type $\boldsymbol{g}=[y,-\partial V/\partial x + k y]^{T}$
dramatically changes the structure of phase curves. While for the pendulum,
the phase curves are closed trajectories around the center, in the case of
pendulum with
friction (constant) they are open spirals converging at the equilibrium after
infinite time. We claim that the pendulum without friction is structurally
unstable. Many dynamicists believe that realistic models of physical processes
should be typical (generic) because we always try to convey the features of
typical garden variety of the dynamical system. The exceptional (non-generic)
cases are treated in principle as less important because they interrupt
discussion and do not arise very often in applications \cite{Abraham:1992}.

In the $2$-dimensional case, the famous Peixoto theorem gives the characterization of the
structurally stable vector field on a compact two dimensional manifold \cite{Peixoto:1962ss}. They
are generic and form open and dense subsets in the ensemble of all dynamical
systems on the plane. If a vector field $f$ is not structurally stable it
belongs to the bifurcation set.

The space of all conservative dark energy models can be equipped with the
structure of the Banach space with the $C^{1}$ metric.

Let $V_{1}$ and $V_{2}$ be two dark energy models. Then $C^{1}$ distance
between them in the ensemble is
\begin{equation}
d(V_{1},V_{2}) = \max \Big\{\sup_{x \in E} | V_{1,x} - V_{2,x}|, \sup_{x \in E}
|V_{1,xx}-V_{2,xx}|\Big\},
\label{eq:9}
\end{equation}
where $E$ is a closed subset of configuration space. Of course, ensemble of
all dynamical systems of Newtonian type on the plane is infinite dimensional
functional space and the introduced metric is a so-called Sobolev metric.

While there is no counterpart of the Peixoto theorem in higher dimensions it can be
easy to test whether planar polynomial systems, like in considered case, have
structurally stable phase portraits. For this aim the analysis of behavior of
trajectories at infinity should be performed. One can simply do that using the
tools of Poincar{\`e} $S^{2}$ construction, namely by projection trajectories
from the center of the unit sphere $S^{2}=\big\{ (X,Y,Z) \in \mathbf{R}^{3} :
X^{2}+Y^{2}+Z^{2}=1 \big\}$ onto the $(x,y)$ plane tangent to $S^{2}$ at either
the north or south pole.

The vector field $f$ is structurally unstable if:
\begin{enumerate}
\item{there are non-hyperbolic critical points on the phase portrait,}
\item{there is a trajectory connecting saddles on the equator of $S^{2}$.}
\end{enumerate}
In opposite cases if additionally the number of critical point and limit cycles
is finite, $f$ is structurally stable on $S^{2}$.

\begin{figure}
\begin{center}
\includegraphics[scale=0.85]{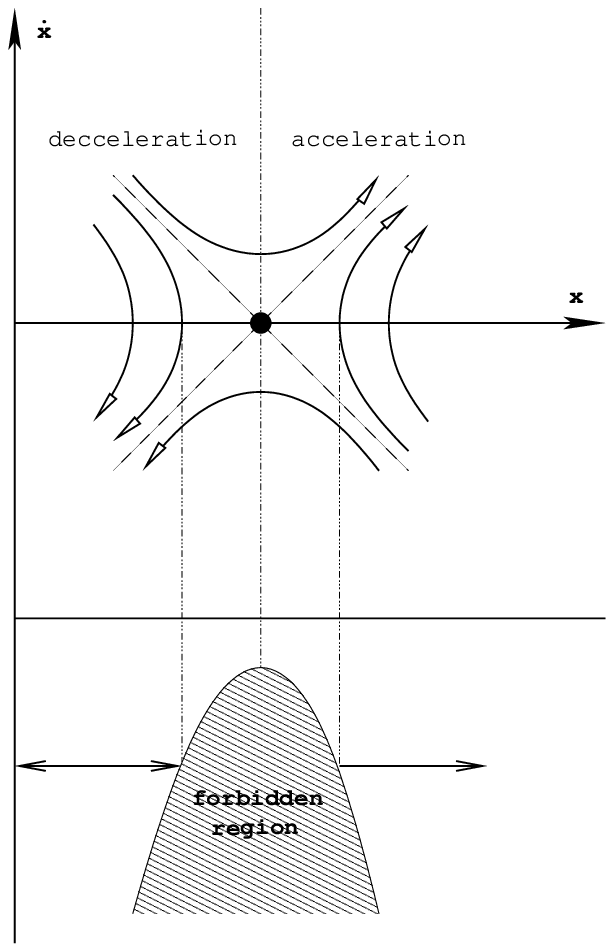}
\includegraphics[scale=0.85]{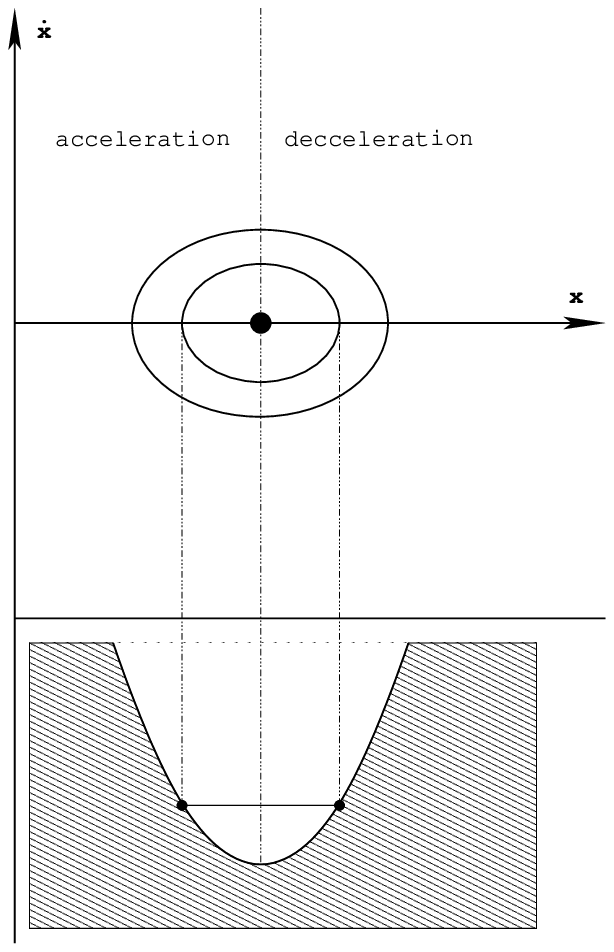}
\includegraphics[scale=0.85]{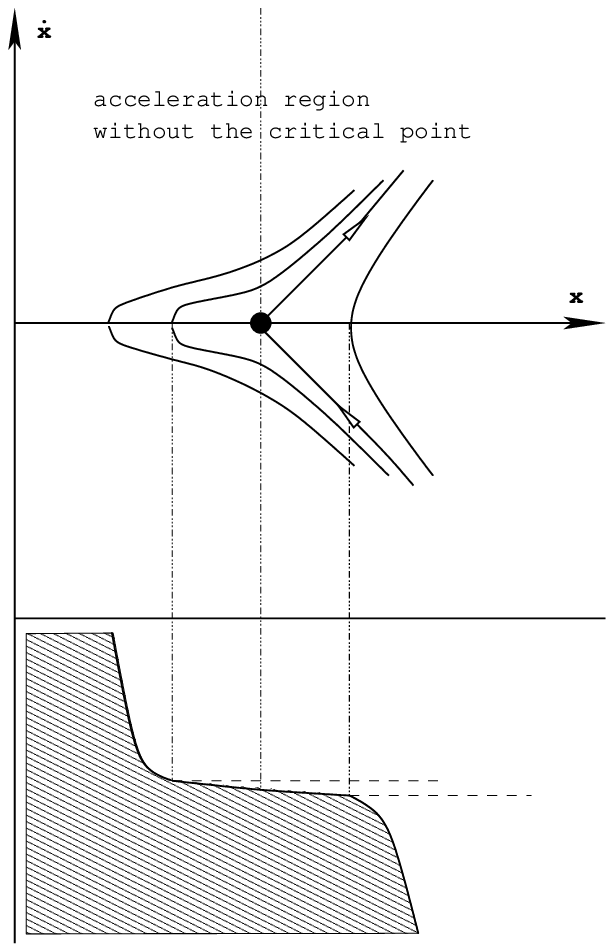}
\end{center}
\caption{Three possible types of behavior in the neighborhood of a critical
point (from left to right): a saddle, a center and a cusp.}
\label{fig:1}
\end{figure}

Let us consider $2$-dimensional dynamical system of a Newtonian type (\ref{eq:5}). There are
three cases of behavior of the system admissible in the neighborhood of the
critical point $(x_{0},0) \colon -\partial V/ \partial x |_{x_{0}}=0$:
\begin{itemize}
\item{If $(x_{0},0)$ is a strict local maximum of $V(x)$, it is a saddle point;}
\item{If $(x_{0},0)$ is a strict local minimum of $V(x)$, it is a center;}
\item{If $(x_{0},0)$ is a horizontal inflection point of $V(x)$, it is a cusp.}
\end{itemize}
All these cases are illustrated in Fig.~\ref{fig:1}.

It is a simple consequence of the fact that characteristic equation for
linearization matrix at the critical point $(\tr{A} = 0 )$ is $\lambda^{2} +
\det{A} = 0$, where $\det{A} = \partial^{2} V/\partial x^{2} |_{x_{0}}$.
Therefore the eigenvalues are real of opposite sign for saddle point, and for
centers -- if
they are non-hyperbolic critical points -- purely imaginary and conjugated.

\begin{figure}
\begin{center}
\includegraphics[scale=1]{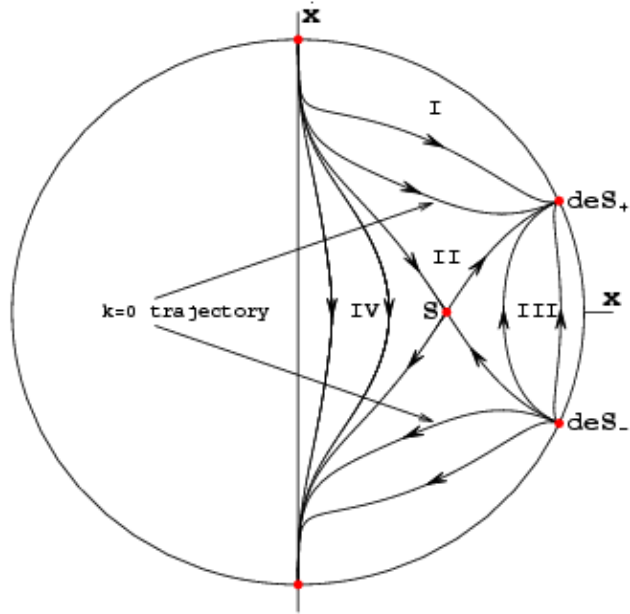}
\end{center}
\caption{The phase portrait for the $\Lambda$CDM model. In the phase portrait
we have a single saddle point in the finite domain and four critical points
located on the circle at infinity. They represent an initial singularity
$(x=0,\dot{x}=\infty)$ or a de Sitter universe (deS). The trajectory of the flat
model ($k=0$) divides all models in to two disjoint classes: closed and open.
The trajectories situated in region II confined by the upper branch of the
$k=0$ trajectory and by the separatrix going to the stable de Sitter node
$\textrm{deS}_{+}$ and by the separatrix going from the initial singularity to
the saddle point $S$ correspond to the closed expanding universe. The
trajectories located in regions I and II (which corresponds to the open
universes) are called inflectional. Quite similarly, the trajectories situated
in region III correspond to the closed universes contracting from the unstable
de Sitter node towards the stable de Sitter node. The trajectories running in
this region describe the closed bouncing universes. The trajectories located in
the region between $\dot{x}$ axis and separatices of the critical point $S$
correspond to the oscillating closed universes expanding from the
initial singularity located at $(x=0,\dot{x}=\infty)$ towards the final
singularity at $(x=0,\dot{x}=-\infty)$.}
\label{fig:2}
\end{figure}

In Fig.~\ref{fig:2} the phase portrait for the $\Lambda$CDM model is presented
on compactified projective plane by circle at infinity. Of course it is
structurally stable. Therefore, following the Peixoto theorem, it is generic in
the ensemble $\mathcal{M}$ of all dark energy models with $2$-dimensional phase space
because such systems form open and dense subsets.

\begin{figure}
\begin{center}
\includegraphics[scale=0.5,angle=270]{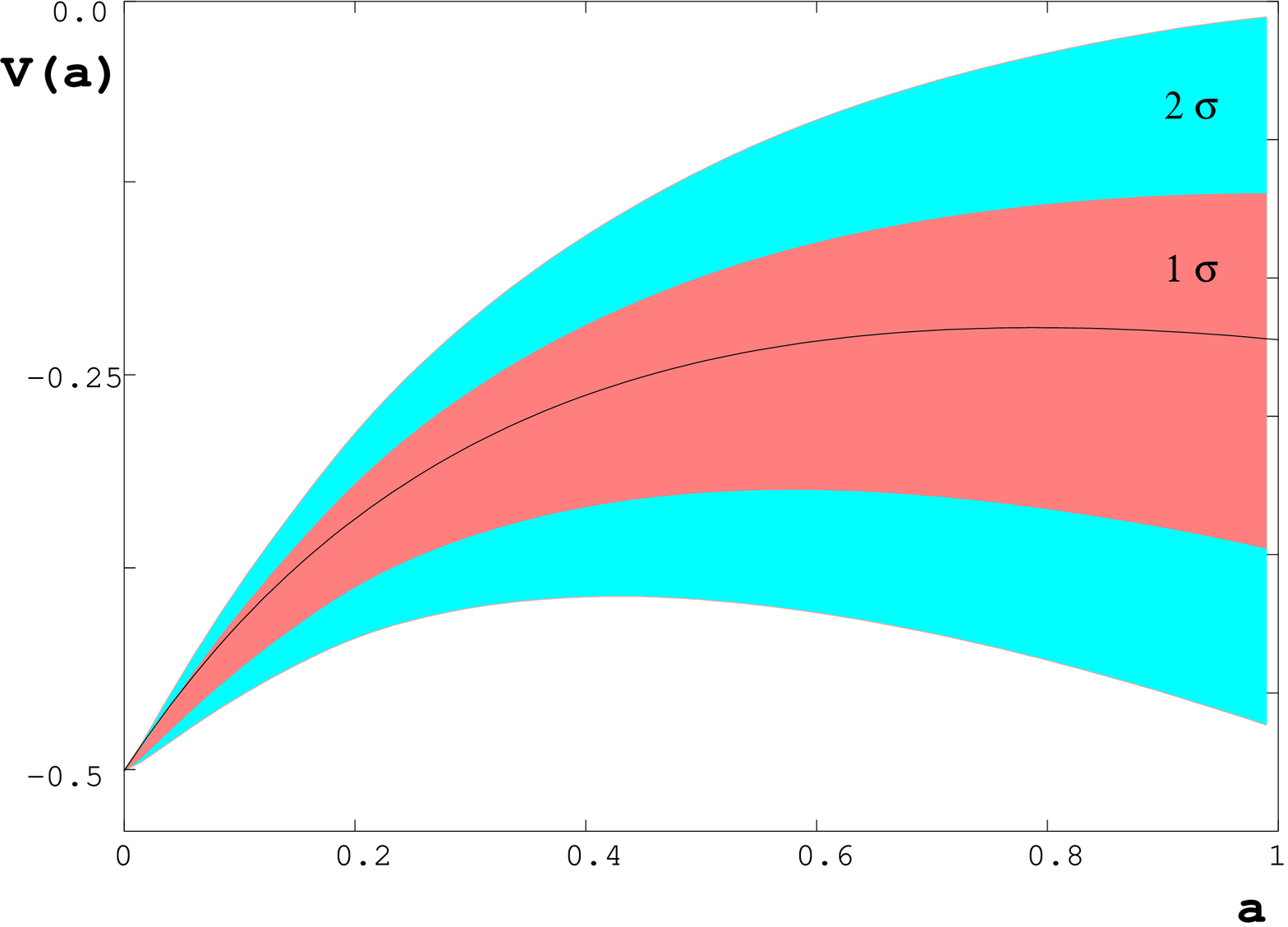}
\end{center}
\caption{The potential function as a function of the scale factor expressed in
its present value $a_{0}=1$ for the reconstructed best fit model is given by
the solid line. The confidence regions $1\sigma$ and $2\sigma$ are drawn around
it. The phase portrait obtained from this potential (best fit) is equivalent to
$\Lambda$CDM (see Fig.~\ref{fig:2}). The value of the redshift transition
estimated from SNIa data (Gold sample) is about $0.38$ (see
\cite{Czaja:2004}).}
\label{fig:3}
\end{figure}

It is also interesting that the phase space of dark energy model can be
reconstructed from SN Ia data set and is topologically
equivalent to the $\Lambda$CDM model. Fig.~\ref{fig:3} represents the potential
function
\begin{equation}
V(a(z)) = -\frac{1}{2}(1+z)^{2} \Bigg[\frac{\ud}{\ud z}
\frac{d_{L}(z)}{1+z}\Bigg]^{-2},
\label{eq:10}
\end{equation}
reconstructed from the relation $d_{L}(z)$ -- the luminosity distance $d_{L}$ as a
function of redshift $z\colon 1+z=a^{-1}$. Such a reconstruction is possible due to
the existence of the universal formula
\begin{equation}
\frac{d_{L}(z)}{1+z} = \int \frac{\ud z'}{H(z')},
\label{eq:11}
\end{equation}
for the flat model.

The main aims of presented discussion is searching for generic cases of
brane-world models which are generic in the ensemble of brane-world models.
We argue that adequate brane models should be located in the near
$\varepsilon$-neighborhood of the $\Lambda$CDM model which is structurally stable.
Therefore the global structure of their dynamics equivalent to the $\Lambda$CDM model is
required. This implies that structural stability becomes requirement which contain
model parameter.

\section{Brane dark energy models as a dynamical systems. Concluding remarks}

Different brane models which offer explanations of acceleration of the current
Universe can be characterized in terms of an additional term
$V_{\textrm{brane}}(a)$ as a function of the scale factor. In Table~\ref{tab:1}
we complete most popular brane models. They might provide answers to a number of
cosmological puzzles including the issue of dark energy or type of the
cosmological singularity. Model called Sahni-Shtanov was introduced in
\cite{Sahni:2003}. There are two classes of such models Brane I and Brane II. Model
called Shtanov-Sahni which describe bouncing model with timelike extra
dimensions was introduced in \cite{Shtanov:2003}.

\begin{figure}
\begin{center}
a)\includegraphics[scale=1]{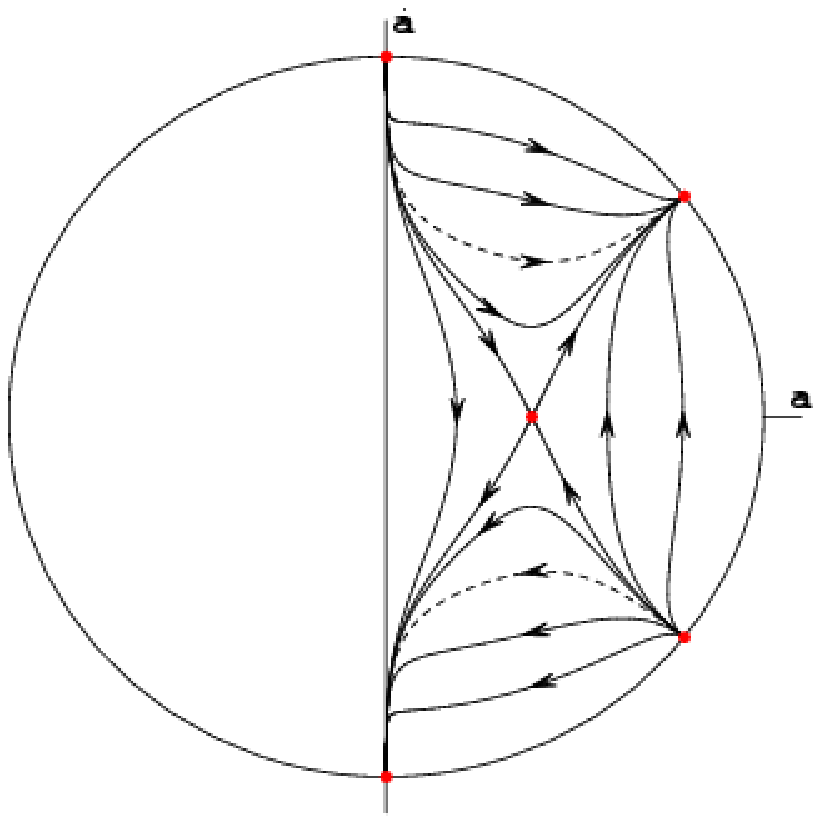}
b)\includegraphics[scale=1]{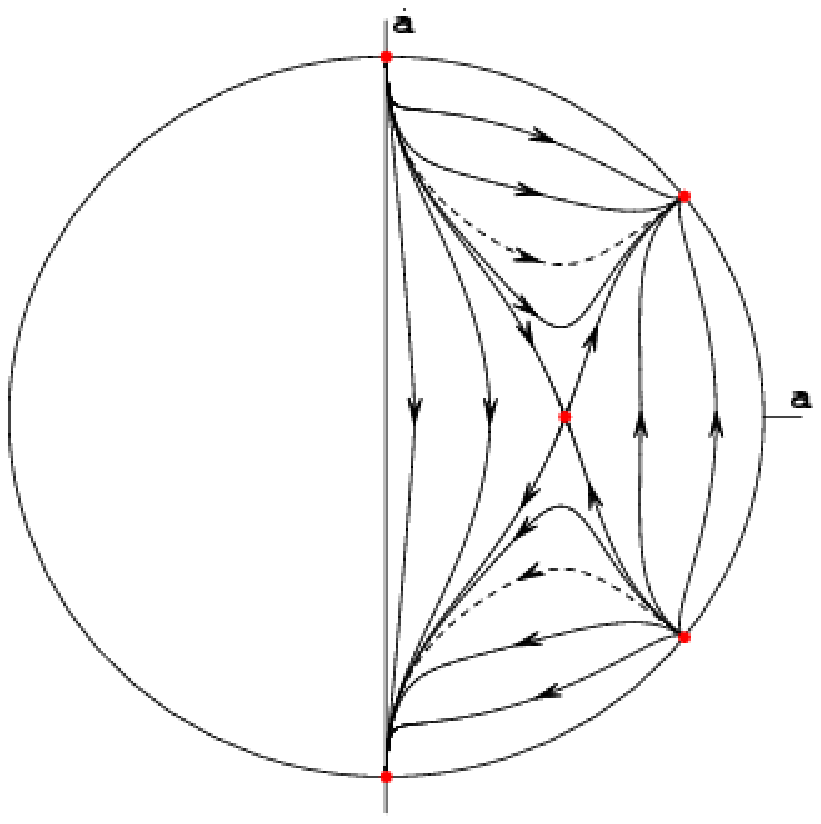}
\end{center}
\caption{Structurally stable phase portraits of Deffayet-Dvali-Gabadadze model model
with $\Omega_{\textrm{m},0}=0.3$, $\Omega_{r_{c}}=0.15$ and: a) $w=0$, b)
$w=1/3$. Dotted trajectories represent flat model trajectories
$\Omega_{k,0}=0$.}
\label{fig:4}
\end{figure}

\begin{figure}
\begin{center}
a)\includegraphics[scale=1]{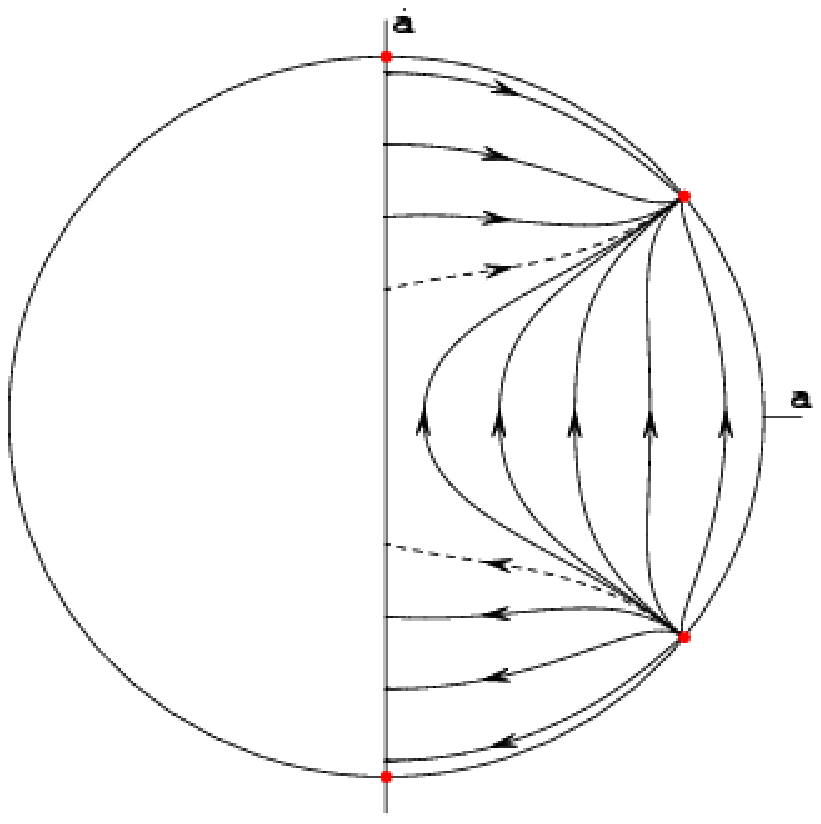}
b)\includegraphics[scale=1]{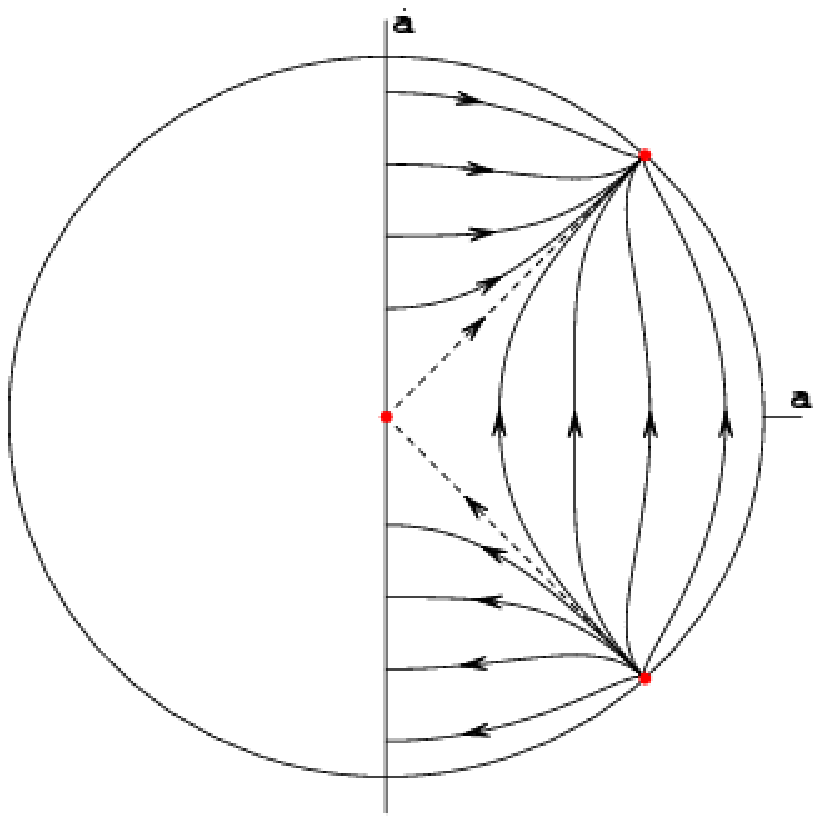}
c)\includegraphics[scale=1]{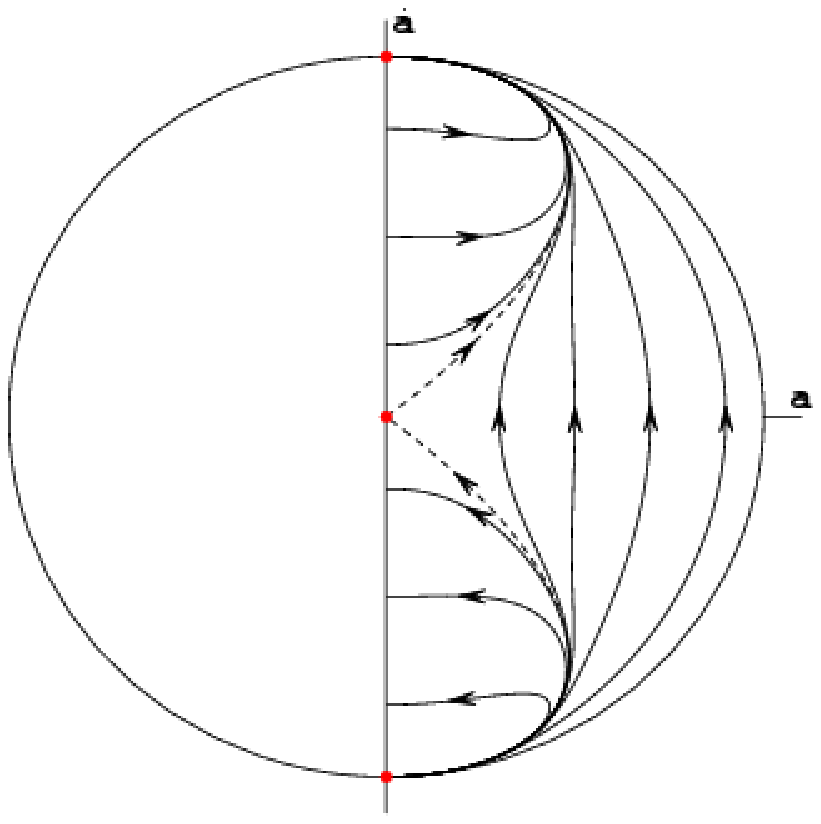}
\end{center}
\caption{Degenerated phase portraits of Deffayet-Dvali-Gabadadze model model with
$\Omega_{\textrm{m},0}=0.3$, $\Omega_{r_{c}}=0.15$ and: a) $w=-1/3$, b) $w=-1$
and c) $w=-4/3$.}
\label{fig:5}
\end{figure}

\begin{figure}
\begin{center}
a)\includegraphics[scale=1]{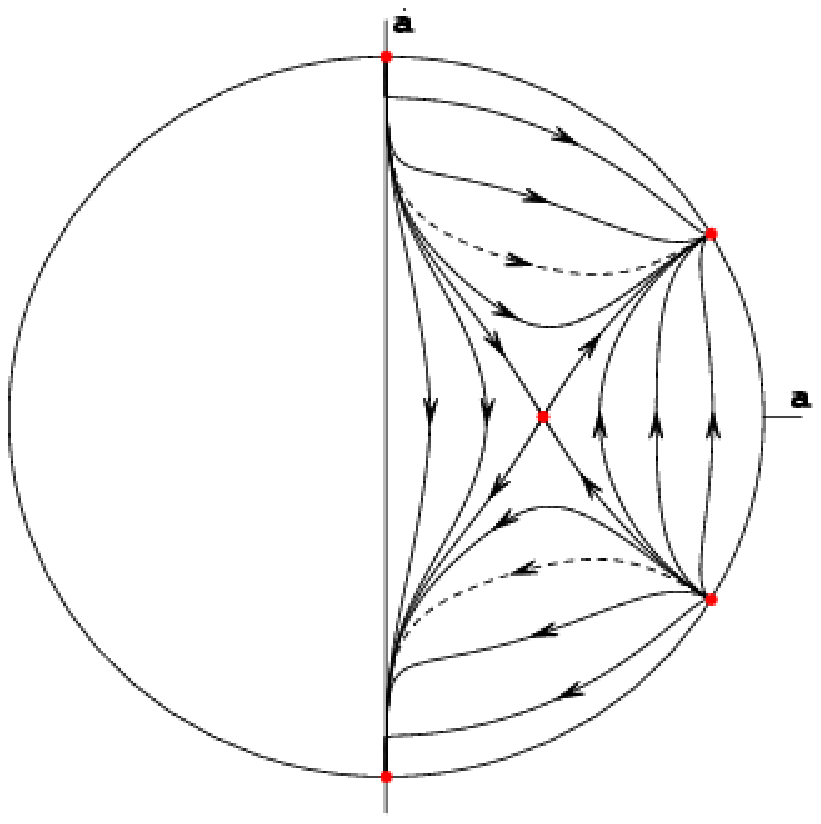}
b)\includegraphics[scale=1]{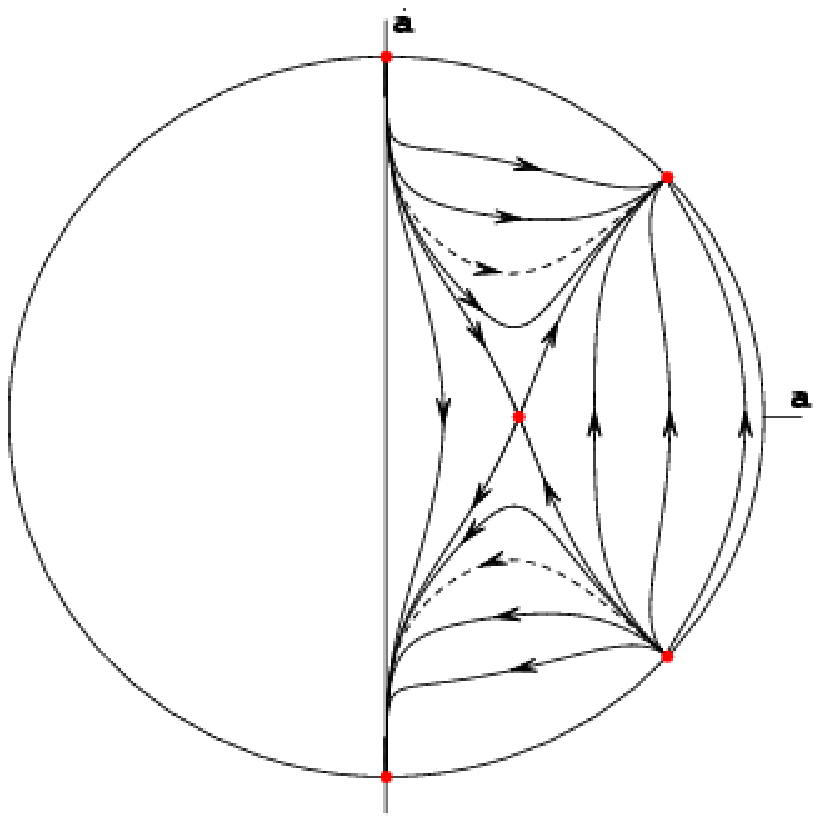}
\end{center}
\caption{Sahni-Shtanov model for $\Omega_{\textrm{m},0}=0.3$,
$\Omega_{\Lambda,0}=0.05$, $\Omega_{l,0}=0.2$ and: a) brane I model (+), b)
brane II model (-).}
\label{fig:6}
\end{figure}

\begin{figure}
\begin{center}
a)\includegraphics[scale=1]{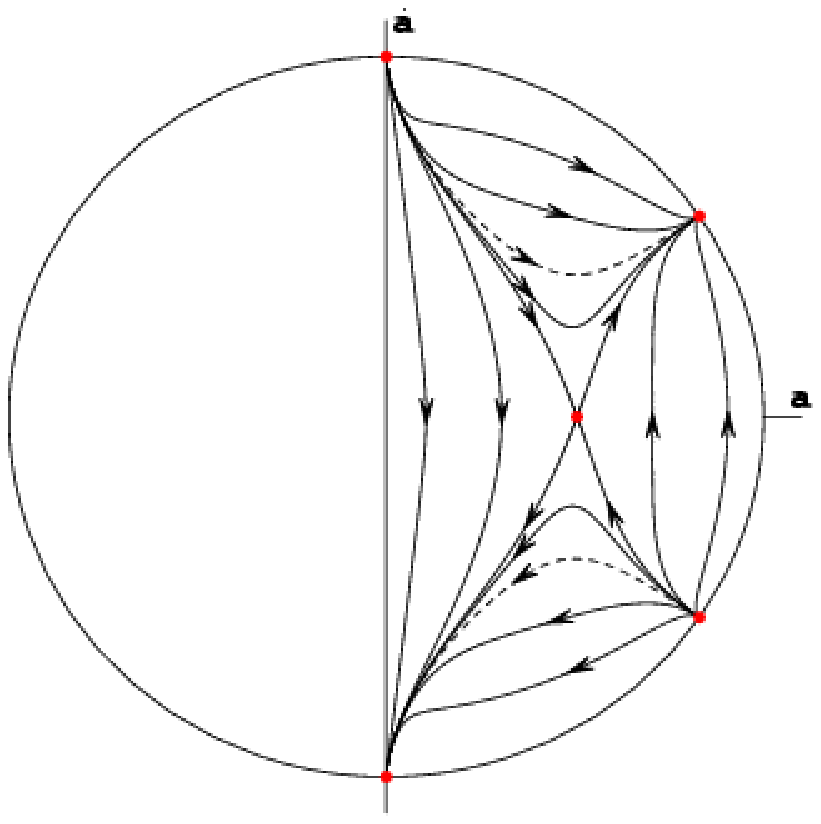}
b)\includegraphics[scale=1]{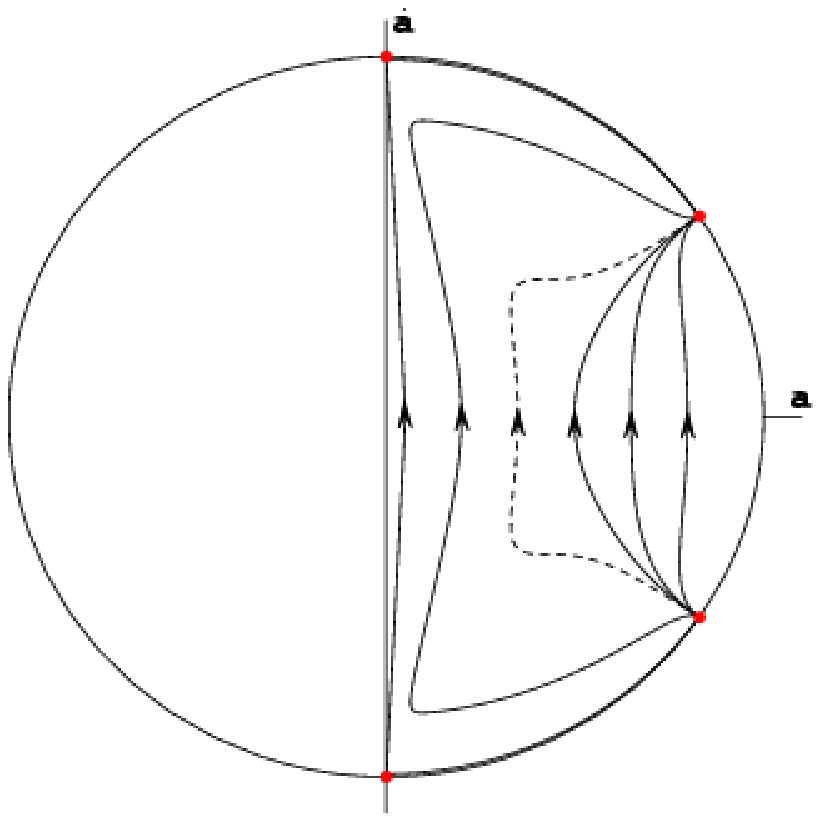}
c)\includegraphics[scale=1]{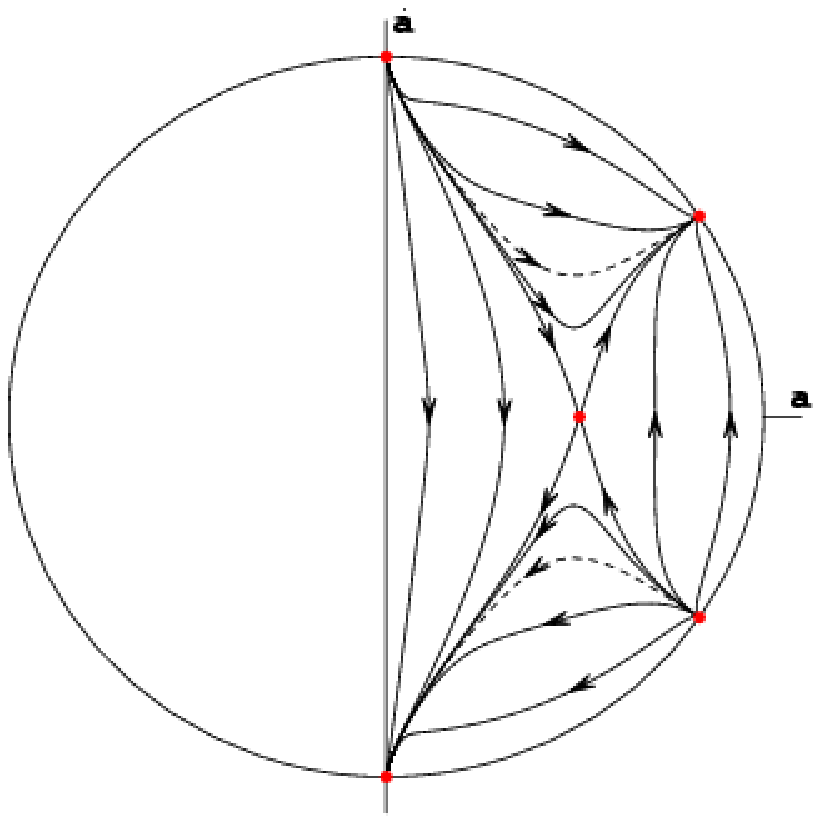}
d)\includegraphics[scale=1]{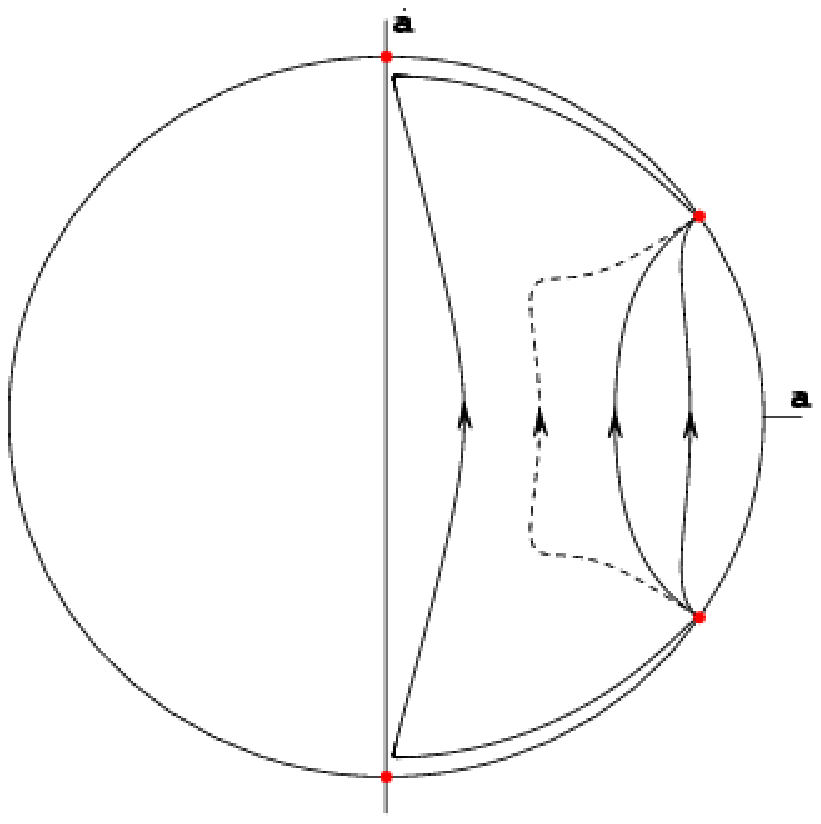}
\end{center}
\caption{Shtanov-Sahni model for $\Omega_{\textrm{m},0}=0.3$, 
$\Omega_{\Lambda,0}=0.45$, $\Omega_{dr,0}=0.15$ and: a) $w=0$, $\epsilon=1$, b)
$w=0$, $\epsilon=-1$, c) $w=1/3$, $\epsilon=1$, d) $w=1/3$, $\epsilon=-1$.}
\label{fig:7}
\end{figure}

\begin{figure}
\begin{center}
a)\includegraphics[scale=1]{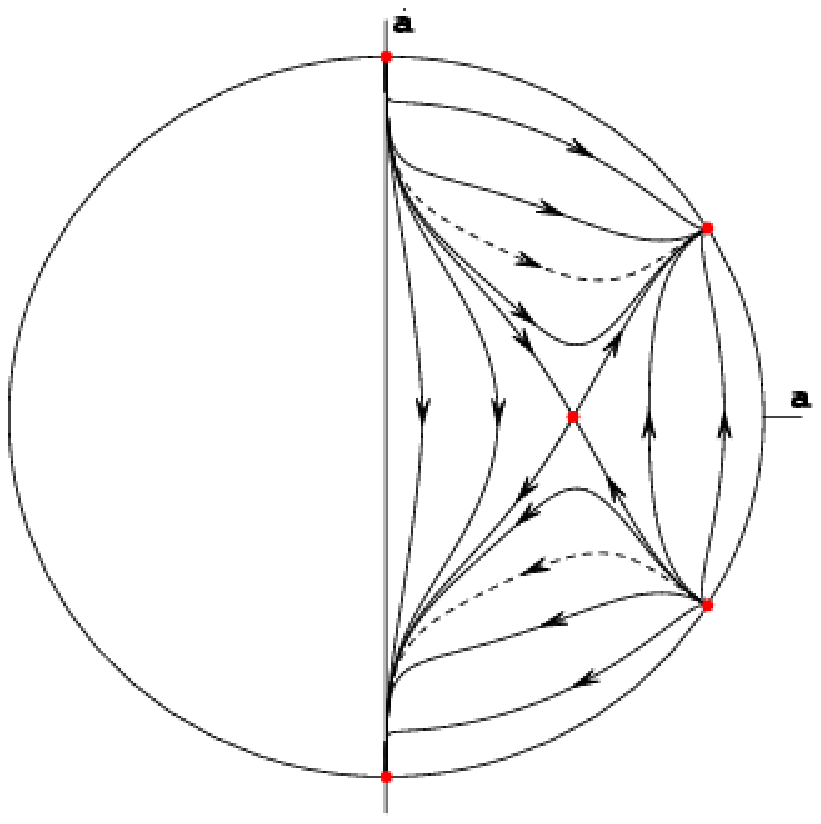}
b)\includegraphics[scale=1]{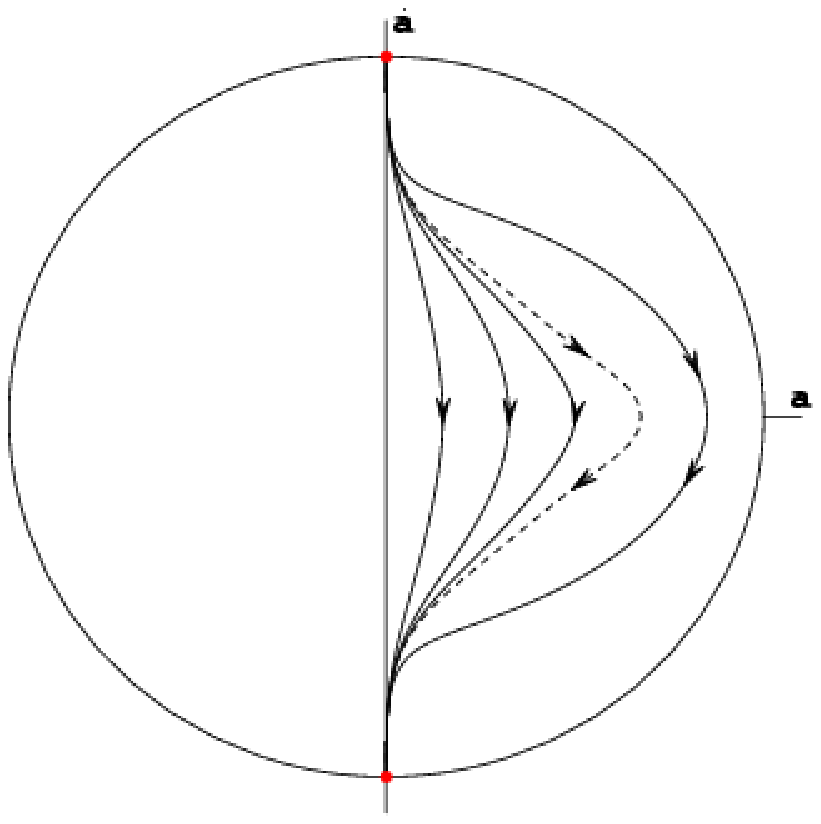}
\end{center}
\caption{Phase portraits for Dvali-Turner model for $\Omega_{\textrm{m},0}=0.45$. From
Table~\ref{tab:1} for $\alpha=1$ we receive a quadratic equation for the variable
$y=(H/H_{0})(a/a_{0})$ and hence two solutions for the potential function: a)
with a plus sign, b) with a minus sign.}
\label{fig:8}
\end{figure}

\begin{figure}
\begin{center}
a)\includegraphics[scale=1]{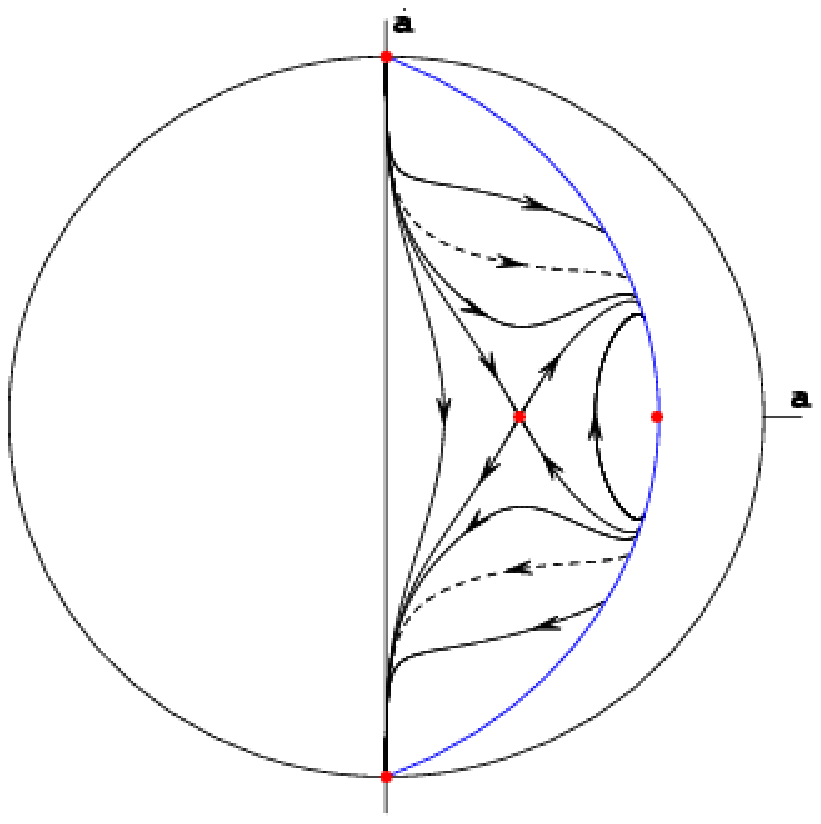}
b)\includegraphics[scale=1]{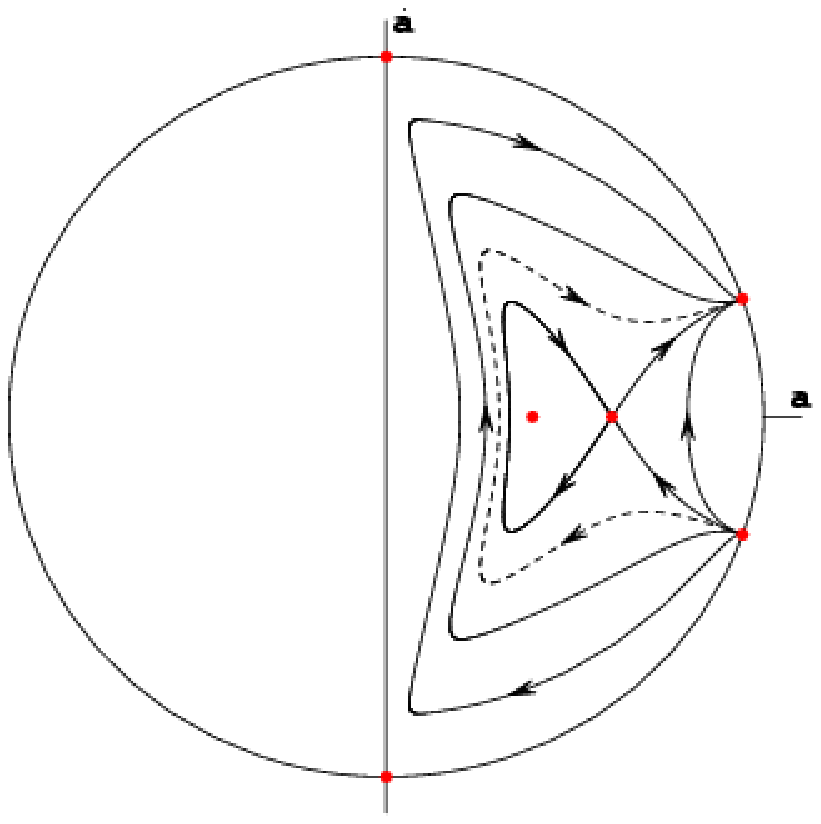}
\end{center}
\caption{a) Phase portrait for Sahni--Shtanov brane I model with
$\Omega_{\textrm{m},0}=0.2$, $\Omega_{\Lambda,0}=0$ and $\Omega_{l,0}=0.32$.
Dashed lines denote flat model trajectories $\Omega_{k,0}=0$. Solid line (blue
in el. version) denotes maximal allowed value of scale factor $a$ for which square root in
case IV (see Table \ref{tab:1}) is positive \cite{Shtanov:2002} . b) Phase portrait for
Shtanov--Sahni model with $\Omega_{\textrm{m},0}=0.3$, $\Omega_{dr,0}=0.47$,
$\Omega_{\Lambda,0}=0.12$ and $w=0$, $\epsilon=-1$. This class of brane--world
models allows for a transient acceleration of the universe which is proceeded
and followed by a matter domination era. Such models admit quiescent
singularities. Model with time-like extra dimension can also avoid cosmological
singularity by a bounce.}
\label{fig:9}
\end{figure}

The Einstein equations on the brane, in general, form a very complicated system
of non-linear partial differential equations. However, the majority of most
interesting models from the cosmological point of view belong to the class of
homogeneous and isotropic ones, for which this complicated system of equations
reduces to the form of a dynamical system. Hence in investigation of dynamics of
these models the methods of dynamical systems seems to be natural. A main
advantage of this methods is that we can obtain on the phase portraits global
visualization of the dynamics, i.e., all evolutional paths admissible for all
initial conditions. Hence we obtain asymptotic states of the system and we can
analyze their type of stability (character of a critical point). The full
knowledge of dynamics require analysis of dynamical behavior not only in a finite
domain of the phase space but also at infinity. In our paper such an analysis was
performed using the technique of Poincar{\'e} compactification of a plane by a circle at
infinity. Although the application of qualitative methods in investigation of
differential equations does not reveal unexpected properties of the brane models
unnoticed until now, this methods offers a possibility of their classification
in the terms of the evolutional paths. Moreover one detect in the phase space
all global attractors and their insets. From the physical point of view it is
interesting to know how large is inset or outset of limit states because the
probability of an initial state of the observation to evolve asymptotically to a
limit set is proportional to the volume of its inset. For example attractors
have open insets, so they are probable (the inset of an attractor is called its
basin).

In our analysis we mainly concentrate on the analysis of brane dynamics and its
reduction to the dynamical system of a Newtonian type. But we also note that some
alternative representation of the dynamics can be useful in the case of Dvali
and Turner model. In some sense in this paper we extend previously introduced
methods \cite{Szydlowski:2006a,Szydlowski:2006b} in large class of brane models.

Apart of detailed analysis of the brane dynamics using dynamical systems
techniques we are looking for the generic dynamical systems of brane origin.
Usually exceptional cases are more complicated from the mathematical point of
view and they interrupt the discussion. Moreover dynamicists believe that they
should not arise very often in the application, because they are exceptional
(\cite[p.349]{Abraham:1992}). Abraham and Shaw pointed out in their famous book
that a considerable portion of the history of the mathematical dynamics has been
dominated by the search for generic properties. The idea was to characterize a class
of phase portraits that are far simpler than arbitrary ones. The main motivation
of the search is to complete classification of the complexity of the phase
portraits (modulo exceptional cases).

This idea was achieved for dynamical systems on the plane by Peixoto due
to fundamental results obtained by Russian mathematicians Andronov and
Leontovich. Peixoto theorem characterize generic dynamical systems on the plane
in tools of notion structural stability which was previously introduced by
Andronov and Leontovich. This theorem states that structurally stable systems on
the $2$-dimensional closed space forms open and dense subsets in the functional space of
the dynamical systems on the plane. The structurally unstable dynamical system
are exceptional in this space.

The main goal of the search of structural stability of dynamical systems of
brane origin on the compactified Poincar{\'e} sphere is to narrow down the
complexity of the portraits enough to allow classification of generic cases.
This was simply achieved because for dynamical systems on the plane with
polynomial right hand sides we have simple test of structural stability
(instability). The notion of structural stability is intuitively very simple: A
vector field has the property of structural stability if all delta perturbations
of it (sufficiently small) have epsilon equivalent phase portraits. The
equivalence of the phase portraits is established by homeomorphism preserving
the orientation of the phase curves (topological equivalence).

It is pointed out that the brane world models have certain features that
distinguish them form the other models of dark energy. They can allow for a
``transient'' universe acceleration phenomena which is preceded and also
followed by matter domination era. Also brane models admit strange type of
singularities which are known as a ``quiescent'' cosmological singularities. It
is also possible that initial singularity can be replaced by characteristic
bounce. The question is: Whether these phenomenas are attributes of a generic
brane dynamical systems? Our answer is No. Typical brane models (on the plane)
have global dynamics represented by phase portraits which should be
topologically equivalent to the $\Lambda$CDM model.

The brane models which a topological structure of the phase space is equivalent to
the $\Lambda$CDM phenomenological model provides a simple alternative to explanation
of accelerating expansion of the current universe. In the concordance
$\Lambda$CDM model parameter $\Lambda$ is constant with respect to redshift $z$
and matter density evolves with redshift like $(1+z)^{3}$. Hence appear basic
problem: Why do they approximately equal each other now ? This problem called
coincidence conundrum is not solved in the frame of the $\Lambda$ model.
Alternatively there is a possibility that there is no dark energy, but instead an
infrared modification of the Friedmann equation on very large scales. To solve this
problem in the frame of the Deffayet-Dvali-Gabadadze (DDG) model it is required $r_{c} \sim H_{0}^{-1}$.

We show that the DDG(+) (Fig. \ref{fig:4}) model has phase portrait equivalent to the $\Lambda$CDM
model which is a purely phenomenological theory offering the description of
acceleration of the current universe rather than its understanding (note that
they does not explain why the vacuum energy does not gravitate
\cite{Maartens:2006}).

The notion of structural stability allows us to formulate some general
assumptions about any adequate brane models offering the explanation of acceleration
of the current universe in terms of the potential function:
\begin{itemize}
\item{the potential function has a maximum at some finite value of the scale factor or
redshift; it is a redshift transition value corresponding to switching a
deceleration phase to an accelerating one,}
\item{at the late time the potential function behaves like $V \propto - a^{2}$ which
guarantee the existence of deS$_{+}$ global attractor.}
\end{itemize}

In our approach the dynamics is reduced to the $2$-dimensional phase plane compactified by
a circle at infinity. If the set of all vector fields $\boldsymbol{f} \in
C^{r}(\mathcal{M}^{2})$ ($r \ge 1$) having a certain property contains an open
dense subsets of $C^{r}(\mathcal{M}^{2})$ then the property is called generic.

In the zoo of brane-world models all discussed previously phenomenological
models can be recovered. In this way we can find counterparts of phantom models
within class of Sahni-Shtanov models as well as the Cardassian one within the
DDG(-) (Fig. \ref{fig:5})
brane models.

Most interesting brane-world models should be, similar to the $\Lambda$CDM model,
structurally stable. Such a postulate is satisfied by the DDG(+) (Fig.
\ref{fig:4}) models as well as
some Sahni-Shtanov models (Fig. \ref{fig:7}) and Dvali-Turner models (Fig.
\ref{fig:8}). In our opinion they should be treated seriously as
candidates for explanation of current SNIa data and other concordance
astronomical observations. The advantage of this models is that they offer some
physical mechanism of this acceleration in contrast to the $\Lambda$CDM model.

From the physical point of view it is interesting to know whether certain
subsets of ensemble contain an open and dense subsets because it means that
this property is typical.

Recently the Bayesian Information Criterion (BIC) was applied in the context of
finding (choosing) adequate accelerating model of the universe
\cite{Davis:2007}. The authors showed preferences for the models beyond
standard FRW cosmology whose best fit parameters reduce them to the cosmological
constant model. We characterize exotic cosmological models in the terms of
structural stability and equivalence of phase portraits.

It was pointed out \cite{Coley:1992} that
concepts of rigidity and fragility formulated in terms of structural stability
may be very useful for cosmology which operates by models. Moreover, as put
forward many years ago by Andronov and Pontryagin \cite{Andronov:1937sg} the structural stability seams
to be important condition for building adequate models of physical phenomenons
in all branches of science. In general, there is a widespread opinion that
physically realistic models of the world should possess some kind of structural
stability. Because, there can by many dramatically different mathematical
models all agreeing with observations and it would be fatal for the empirical
methods of science \cite{Thom:1977} if we are taking into account final errors of
measurements \cite{Szydlowski:1984}. In
our opinion structural stability of dark energy models (note that $\Lambda$CDM
is favored by astronomical data by Bayesian selection framework
\cite{Szydlowski:2006ay, Szydlowski:2006pz, Kurek:2007tb, Kurek:2007gr})
reflects their flexibility with respect the data fitting.

We always try to convey the features of typical garden variety of dynamical
system accelerating cosmological models. The exceptional cases are usually more
complicated and they in principle interrupt the discussion. This prejudice is
shared by all dynamicists (\cite[p.149]{Abraham:1992}). We find that brane
models in which instead of initial singularity there is present bounce are
structurally unstable because of the center on the phase portraits. Therefore
the bounce is not generic property of the evolutional scenario of accelerating
brane models. Such a models require presence of double accelerating phases
during the cosmological evolution. In terms of potential function of dynamical
system it means that the existence of minimum before the maximum. If we consider
the Sahni-Shtanov brane II model then for some set of parameters appears a quiescent
singularity at which the density, pressure and the Hubble function remain finite,
while all the Riemannian invariants diverge to infinity. Similar models belong to
the class of non-generic cosmological models. Note that all models within the
accelerating phase, which is only a transitional phenomenon, are non-generic too. As it
was demonstrated by Sahni and Shtanov \cite{Sahni:2004} there is a class of
loitering brane-world models. In the phase portraits the corresponding
evolutional paths are situated near the separatrices of saddle point
(inflectional models). This property is generic in contrast to
a quiescent future singularity. 

\begin{acknowledgments}
The authors are very grateful to Youri Shtanov for comments and suggestions.
This work has been supported by the Marie Curie Host Fellowships for the
Transfer of Knowledge project COCOS (Contract No. MTKD-CT-2004-517186).
\end{acknowledgments}

\end{document}